\documentclass[aps,prd,twocolumn,showpacs,superscriptaddress,groupedaddress]{revtex4-1}
\usepackage{graphicx}
\usepackage{dcolumn}
\usepackage{bm}
\usepackage{amssymb}
\usepackage{amsmath}

\begin{document}

\title{Neutrino Production in Electromagnetic Cascades: An extra component of cosmogenic neutrino at ultrahigh energies}

\author{Kai Wang$^{1,2}$}\email{wang@mpi-hd.mpg.de}
\author{Ruo-Yu Liu$^2$}\email{ruoyu@mpi-hd.mpg.de}
\author{Zhuo Li$^{4,5}$}
\author{Zi-Gao Dai$^{1,3}$}

\affiliation{$^1$School of Astronomy and Space Science, Nanjing University, Nanjing 210093, China\\
$^2$Max-Planck-Institut f\"ur Kernphysik, 69117 Heidelberg, Germany\\
$^3$Key Laboratory of Modern Astronomy and Astrophysics (Nanjing University), Ministry of Education, Nanjing 210093, China\\
$^4$Department of Astronomy, School of Physics, Peking University, Beijing 100871, China\\
$^5$Kavli Institute for Astronomy and Astrophysics, Peking University, Beijing 100871, China}

\date{\today}

\begin{abstract}
Muon pairs can be produced in the annihilation of ultra-high energy (UHE, $E \gtrsim 10^{18} \,\mathrm{eV}$) photons with low energy cosmic background radiation in the intergalactic space, giving birth to neutrinos. Although the branching ratio of muon pair production is low, products of other channels, which are mainly electron/positron pairs, will probably transfer most of their energies into the new generated UHE photon in the subsequent interaction with the cosmic background radiation via Compton scattering in deep Klein-Nishina regime. The regeneration of these new UHE photons then provides a second chance to produce the muon pairs, enhancing the neutrino flux. We investigate the neutrino production in the propagation of UHE photons in the intergalactic space at different redshifts, considering various competing processes such as pair production, double pair production for UHE photons, and triplet production and synchrotron radiation for UHE electrons. Following the analytic method raised by Gould and Rephaeli (1978), we firstly study the electromagnetic cascade initiated by an UHE photon, with paying particular attention to the leading particle in the cascade process. Regarding the least energetic outgoing particles as energy loss, we obtain the effective penetration length of the leading particle, as well as energy loss rate including the neutrino emission rate in the cascade process. Finally, we find that an extra component of UHE neutrinos will arise from the propagation of UHE cosmic rays due to the generated UHE photons and electron/positrons. However, the flux of this component is quite small, with a flux of at most $10\%$ of that of the conventional cosmogenic neutrino at a few EeV, in the absence of a strong intergalactic magnetic field and a strong cosmic radio background. The precise contribution of extra component depends on several factors, e.g., cosmic radio background, intergalactic magnetic field, the spectrum of proton, which will be discussed in this work.
\end{abstract}

\maketitle
\section{Introduction}

Cosmogenic neutrinos are expected to be produced in the interaction between ultrahigh energy cosmic rays (UHECRs) and cosmic background radiation in their propagation via the Greisen-Zatsepin-Kuzmin (GZK) mechanism \cite{GZK66}. A fraction of UHECRs energy is transferred into charged pions which will decay and generate neutrinos. Neutral pions, which are also produced associately with charged pions, could be another origin of cosmogenic neutrinos. As these neutral pions decay into gamma rays, generated gamma rays can interact with the soft photons of the cosmic background radiation, and initiate a electromagnetic (EM) cascade \cite{Coppi97,murase09}. Because of the high incident photon energy, a small fraction of the interactions is able to enter the channel of muon pair production (MPP, $\gamma\gamma \to \mu^+\mu^-$), and the produced muons will further decay into neutrinos \cite{li07}. Although an UHE photon will most likely give birth to electron/positron pairs via electron/positron pair production (EPP, $\gamma\gamma \to e^+e^-$) \cite{murase09,murase12,gould67a}, regeneration of a new UHE photon is possible as one of the generated electron and positron will carry most energy of the UHE photon (namely, the leading particle), and it may pass the energy to one of the background photons via Compton scattering (CS) in deep Klein-Nishina (K-N) regime. As such, the energy of the initial UHE photon is conserved in the leading particle and get a second chance to generate muon pairs when the regenerated UHE photon interacts with the cosmic background radiation again. This cycle will proceed many times until the energy of the regenerated photon falls below the threshold of MPP,  largely enhancing the neutrino production. For the first time, Ref.\cite{aharonian92} treat the leading electron or photon as an $e/\gamma$ particle to study its penetration length because this UHE particle spends its lifetime in two states ($e$ or $\gamma$) and most of energy of initial particle is conserved in the leading particle. As shown in some past works (e.g., \cite{gould78} and \cite{bahtt00}), the effective penetration length of UHE photon is $\sim 50-150 \,\mathrm{Mpc}$ at $10^{19}-10^{19.5} \,\mathrm{eV}$, and during this period, during $\sim 1/3$ of the lifetime, UHE particle propagates as a photon. In the cascades, a brief analytical estimation of ratio of UHE photons going to the MPP channel can be up to $10\%$ by comparing the effective penetration length of UHE photons propagating as a photon with the mean free path of the MPP (see Fig.~\ref{f1}) roughly, as long as the energy of UHE photon is smaller than $7 \times  10^{20}\,\mathrm{eV}$ (beyond this energy, double pair production would take over the propagation, see below).

 A strict treatment of propagation of UHE photon in cosmic background radiation has been given in Ref.\cite{aharonian90} based on the kinetic equations for electrons and photons, and the UHE photons produced in transient sources and propagating in the structured region have been discussed in Ref.\cite{murase12}. To calculate this extra component of cosmogenic neutrinos, we need to properly treat the  transformation chain of the leading particle as $\gamma \to e \to \gamma \to e \to \cdot\cdot\cdot$ in the EM cascade initiated by an UHE photon. Ref.\cite{gould78} (hereafter, GR78) used an analytical method to calculate the effective penetration distance of an UHE photon with considering the EPP and the successive CS process, which has been applied in Ref.\cite{aharonian92}. By regarding the energies taken away by the least energetic outgoing particles in the interactions as the energy loss of the leading particle, GR78 obtained the average energy loss rate of the UHE electromagnetic particle, which is either in an electron/positron state or in a photon state.
 In this paper, we will follow the method of GR78 to calculate the energy loss in one cycle of $\gamma \to e \to \gamma$, taking some other processes into consideration, such as double pair production (DPP, $\gamma\gamma \to e^+e^-e^+e^-$) and muon pair production (MPP, $\gamma\gamma \to \mu^+\mu^-$) for the transformation  $\gamma \to e$, and triplet production (TPP, $e\gamma \to e e^+e^-$) and synchrotron for the transformation $e \to \gamma$. In the meanwhile, we can obtain the average emission rate of neutrinos.

While cosmic microwave background (CMB) at different redshifts are well known, cosmic radio background (CRB)  and intergalactic magnetic field (IGMF) are not clear so far,  which, however, are important to the propagation of UHE photons and electrons. CRB photons have lower energies than CMB photons. When they interact with UHE photons or electrons, the leading particle effect is not as pronounced as that in the interaction with CMB photons due to a smaller center--of--mass energy. So the energy of the leading particle will be smaller in the presence of CRB. On the other hand,   synchrotron radiation may also take away the energy of electrons if the intergalactic magnetic field is too strong and stop the regeneration of a new UHE photon. The effect of these two factors on the propagation will be studied in this work.

The rest part of the paper is organized as follows. In Section \ref{sec:epd}, we study the energy loss process of the leading particle in the EM cascade, either as a photon or an electron, and obtain the effective penetration length of initial UHE photon.  We calculate the extra neutrino production via the MPP channel in the propagation of UHE protons in Section \ref{sec:neupro}. Discussions and conclusions are given in Section \ref{sec:d&c}. Details of interactions for UHE photons and UHE electrons considered in this paper can be found in Appendix \ref{photonsection} and Appendix \ref{electronsection} respectively.

\section{The Effective Penetration Distance of Ultrahigh Energy Photons}\label{sec:epd}
\subsection{An analytical Method to estimate the effective penetration length}
Electromagnetic cascades in the deep K-N regime has been widely discussed in the literature \cite{zdzi88+,bahtt00,murase12}. Here, following GR78, we provide an analytical method by regarding the cycle $\gamma \to e \to \gamma$ as an unit process of the transformation chain $\gamma \to e \to \gamma \to e \to \cdot\cdot\cdot$ initiated by an UHE photon. By calculating the energy loss in one cycle, we can get the average energy loss rate of the photon and then obtain its effective penetration distance. Let's first consider the interaction of the UHE particles (photons and electrons) with CMB.

The basic process in the first half cycle $\gamma \to e $ is EPP ($\gamma \gamma \to e^+ e^-$) between the UHE photon and background photons. Denoting the energy of the UHE photon and the background photon respectively by $E_\gamma$ and $\varepsilon$ in unit of the rest energy of electron $m_ec^2$,  most of the UHE photon's energy will be likely carried by a leading particle which is one of the generated electron and positron, as long as $\omega_\gamma \equiv \varepsilon E_\gamma  \gg 1$. Although the other generated particle may be still energetic enough to initiate a cascade, its final products are less important than those produced by the leading particle either by total energy or by number of the generated secondaries. So we will simply regard it as an energy loss following the treatment of GR78. The cross section of the EPP process decreases with the increase of $\omega_\gamma$. Given the typical energy of CMB photon $\epsilon$ to be $2.3\times 10^{-4}(1+z)\,$eV at redshift $z$, the DPP ($\gamma \gamma \to 2e^+ 2e^-$) becomes increasingly important and dominates EPP if the energy of UHE photons reaches $\sim 7\times 10^{20}/(1+z) \mathrm{eV}$, given a constant cross section of DPP $\sigma_{dpp} = 6.45 \times 10^{-30} \,\mathrm{cm^{2}}$  above its threshold \cite{brown73a}. One of the two generated pairs in DPP shares most of the energy of the UHE photon so evenly that there is no distinct leading particle from this channel and hence energies go into this channel will be regarded as energy loss \cite{demidov09}. Another process we consider here for an UHE photon is the muon pair production (MPP, $\gamma\gamma \to \mu^+ \mu^-$). Although the peak cross section of MPP is $(m_e/m_\mu)^2$ times smaller than that of the EPP, the cross section of MPP peaks at much higher energy, due to a much larger threshold energy, where the cross section of EPP already decreases to about only order of magnitude higher than the peak cross section of MPP. Thus, at UHE energies, MPP is a non-negligible energy loss process.

Due to much smaller cross sections, other processes about $\gamma\gamma$ reactions tend to be negligible \cite{brown73b}, e.g., (1) the asymptotic cross sections of $\gamma\gamma \to e^+ e^- \mu^+ \mu^-$ and $\gamma \gamma \to 2\mu^+ 2\mu^-$ are $\sim 4 \times 10^{-4} \sigma_{dpp}$ and $(m_e/m_\mu)^2 \sigma_{dpp} \sim 2\times 10^{-5} \sigma_{dpp}$, respectively; (2) For the reactions $\gamma\gamma \to $hadrons, the cross section would be smaller and the threshold energy would be larger than MPP's, because of its larger mass of hadrons than muons'; (3) For some weak interactions (e.g., $\gamma\gamma \to \nu{\bar \nu }$), the cross sections (typically, $\sim $fb-pb, \cite{senol12}) are too small to be important.

Then, we move to the latter half cycle $e \to \gamma$ which is initiated by the leading electron/positron (hereafter we do not distinguish electron from positron and use the term "electron" for either of them) generated in the EPP. The basic interaction process is the (inverse) Compton scattering. An UHE electron may pass most of its energy to a background photon as long as the scattering occurs in deep K-N regime, i.e., $\omega_e\equiv \varepsilon E_e \gg 1$, while the new generated UHE photon will then start a new cycle of $\gamma \to e \to \gamma$.   However, before the regeneration of the new UHE photon, the UHE electron may undergo some other energy loss processes. While the cross section of CS goes roughly as $1/E_e$,  the cross section of TPP ($e\gamma  \to e e^+ e^-$) increases logarithmically with energy, and becomes comparable to the CS cross section at $\omega_e \simeq 220$ (see Appendix \ref{tppsection}). One of the produced electrons in TPP carries most of the UHE electron's energy, so TPP will not stop the cycle immediately as this leading electron still has a chance to generate an UHE photon via CS. However, the energy taken by the other two less energetic electrons would add up in the subsequent TPPs and make a significant energy loss if the UHE electron undergoes too many times TPP before regenerating a new UHE photon via CS. We regard TPP as a continuous energy loss process, similar to the synchrotron loss in IGMFs, which is another continuous energy loss process we consider here. In addition, there is a small possibility in CS to generate a leading electron rather than a leading photon. We treat these two cases of CS separately in the following calculation, and denote the case of producing a leading electron as ``CS1'' and the case of producing a leading photon as ``CS2''. CS1 is also treated as a continuous energy loss process.

We neglect other less important processes such as the double Compton scattering (\cite{ram71}, $e\gamma  \to e \gamma \gamma$), muon electron-pair production (\cite{ath01}, $e \gamma \to e \mu^+ \mu^-$). The contribution of the former one could be canceled by a radiative correction to single Compton scattering (A comprehensive discussion of this problem would be beyond the scope of this paper. For details, see Ref.\cite{mandl52+}) and the cross section of the latter one is typically $4 \times 10^{-3} \,\mathrm{\mu b} - 0.1 \,\mathrm{\mu b}$ for $s=4 m_\mu^2-20m_\mu^2$.

In summary, we can divide the cycle to two steps: the first step ($\gamma \to e $) is for UHE photons including EPP, DPP, and MPP; the second step ($ e \to \gamma$) is for UHE electrons including CS, TPP and synchrotron radiation. We can write the average energy loss rate in the first step as
\begin{equation}
{\left( {\frac{{dE}}{{dt}}} \right)_{\gamma  \to e}} = {\left( {\frac{{dE}}{{dt}}} \right)_{epp}} + {\left( {\frac{{dE}}{{dt}}} \right)_{dpp}} + {\left( {\frac{{dE}}{{dt}}} \right)_{mpp}},
\end{equation}
while for the second step, the average energy loss rate is
\begin{equation}
{\left( {\frac{{dE}}{{dt}}} \right)_{e \to \gamma }} = {\left( {\frac{{dE}}{{dt}}} \right)_{cs}}+ {\left( {\frac{{dE}}{{dt}}} \right)_{tpp}} + {\left( {\frac{{dE}}{{dt}}} \right)_{syn}},
\end{equation}
with $(dE/dt)_i$ being the corresponding energy loss rate of $i$ process. We denote the interaction probability per unit distance traveled by the UHE particles by $(d\tau/dx)_i$. For those processes in which no leading particle is generated (i.e., none of the produced secondaries is a  photon or electron with energy larger than $E/2$, such as the MPP and DPP), we have the relation $(dE/dt)_i = cE (d\tau/dx)_i$. Once the interaction enters these channels, the cycle will be ceased. By contrast, the cycle will continue as long as a leading particle is generated unless the energy of the leading particle is already lower than half of the initial energy. In these processes such as  EPP, CS and TPP, the relation between $(dE/dt)_i $ and $(d\tau/dx)_i$ would be different from MPP's and DPP's. Derivations of $(dE/dt)_i $ and $(d\tau/dx)_i$ for all the considered processes can be found in the Appendix.


Now, let us consider a group of $N$ UHE photons with identical energy $E$. After the first step, we assume a  fraction $y$ of UHE photons produce UHE electrons via EPP, with an average electron energy $E'=\int_{A/2}^A {{E_e}\frac{{dN}}{{d{E_e}dt}}d{E_e}} /\int_{{E_{e,\min }}}^{A} {\frac{{dN}}{{d{E_e}dt}}d{E_e}} $, where $A=E_\gamma+\epsilon  \approx E_\gamma$. $y$ can be also interpreted as the probability of entering the EPP channel in the first step, and can be found by $y={(d\tau /dx)_{epp}}/\left[ {(d\tau /dx)_{epp}} \right.$ $\left. +{(d\tau /dx)_{dpp}}+{(d\tau /dx)_{mpp}} \right]$. The total energy loss in one cycle $\gamma \to e \to \gamma$ is then
\begin{equation}
\Delta {E_{tot}} = {t_1}{\left( {\frac{{dE}}{{dt}}} \right)_{\gamma  \to e}} + y{t_2}{\left( {\frac{{dE'}}{{dt}}} \right)_{e \to \gamma }},
\end{equation}
and the average energy loss rate can be written as
\begin{equation}\label{dedt}
{\left\langle {\frac{{dE}}{{dt}}} \right\rangle_{tot}} = \frac{{{t_1}}}{{{t_1} + {t_2}}}{\left( {\frac{{dE}}{{dt}}} \right)_{\gamma  \to e}} + {y}\frac{{{t_2}}}{{{t_1} + {t_2}}}{\left( {\frac{{dE'}}{{dt}}} \right)_{e \to \gamma }},
\end{equation}
where $t_1=\left\{ {c\left[ {(d\tau /dx)_{epp}}+{(d\tau /dx)_{dpp}}+{(d\tau /dx)_{mpp}} \right]} \right\}^{-1}$ and $t_2=\left\{ {c[ {(d\tau /dx)_{cs,2}}+ {(dE' /dt)_{cs,1}}/cE'} \right.$ $  \left. {+ {(dE' /dt)_{tpp}}/cE'+{(dE' /dt)_{syn}}/cE'  ]} \right\} ^{-1}$ are the average duration of the first step and the second step of the cycle respectively. Then, the effective penetration distance can be given by
\begin{equation}\label{epd}
{\lambda _{eff}} =  - cE/{\left\langle {dE/dt} \right\rangle _{tot}}.
\end{equation}
We should note that Eq.~(\ref{epd}) is valid to calculate the effective penetration distance only when at least one cycle $\gamma \to e \to \gamma$ can be completed before the total energy of particles decreases dramatically. More specifically, it requires that the DPP process is not important, which roughly translates to $E < 7\times 10^{20}/(1+z) \mathrm{eV}$. On the other hand, when DPP is important, most of energy of the photons will be consumed in the first step of the cycle ($\gamma \to e$) without generating leading particles. In this case, the effective penetration distance is equal to the mean free length of DPP process approximately.

\subsection{Effective penetration length in various environments}
As we briefly mentioned in the Introduction, the cascade process are affected by various factors of the environments, such as the cosmic radio background and the intergalactic magnetic field.

When considering interactions at extremely high energy (EHE, $E>10^{19}\,$eV), photons from CRB can take an important role. For the first half cycle, CRB photons are able to trigger the EPP with EHE photons, while for the latter half cycle, the mean free length of CS for EHE electrons with CMB photons becomes larger than that with CRB photons due to a stronger K-N effect in the former interaction. One would expect a shorter effective penetration distance, because CRB leads to lots of interactions with $\omega_e$ and $\omega_\gamma\gtrsim 1$ in which the leading particle effects in EPP and CS is not evident.
Due to the large uncertainty of CRB, we consider two cases in the first one of which CRB is ignored, while we adopt the CRB estimated by \cite{pro96} including normal galaxies and radio galaxies with no evolution as redshifts in the second case.

We note that the strength of IGMF is not clear so far, which, could  be $nG-\mu G$ in the structured region (clusters and filaments) and could be much smaller in voids \cite{gov04+,ner09}. Thus, we also set the strength of IGMF as a free parameter to study the effect of synchrotron. When considering a strong IGMF, synchrotron loss of electrons would be more important than other processes of electrons beyond certain energy, so that the transformation $e \to \gamma$ in the latter half cycle will be suppressed. The stronger is the magnetic field, the smaller is the energy from which synchrotron loss starts to have influence.

 \begin{figure}[hptb]
\includegraphics[width=1.0\linewidth]{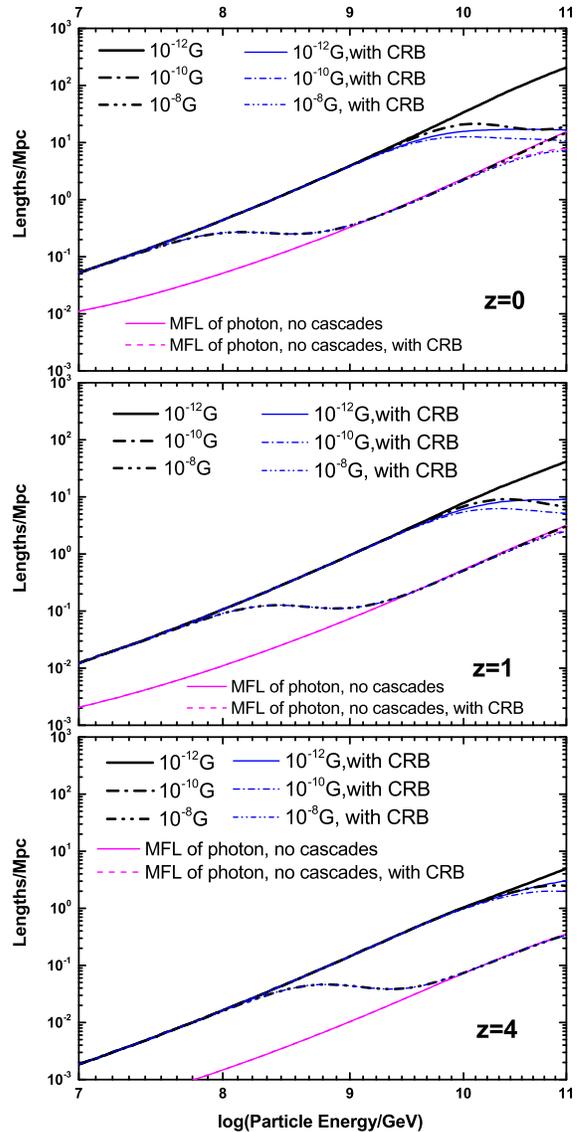}
\caption{\small{\label{f3}
The effective penetration distances for a initial UHE photons at $z=0$, $z=1$ and $z=4$. The black thick lines represent the effective penetration distances without CRB in different magnetic fields, while the blue thin lines present the effective penetration distances with CRB in different magnetic fields. The magenta thin solid and dashed lines represent the mean free lengths of photons in the cases without CRB and with CRB, respectively.
}}
\end{figure}

We exhibit the effective penetration distances v.s. photon initial energy in Fig.~\ref{f3}.
The top, middle, bottom panel are for $z=0$, $z=1$ and $z=4$ respectively. Black (blue) lines show the effective penetration distance without (with) CRB, while different types of lines represent the results with different strength of IGMF. Pink solid (dashed) lines are the mean free lengths of a photon without (with) considering CRB for comparison. Take the top panel for example, the effective penetration distance of a photon is $\sim 10$ times larger than its mean free length in the absence of the CRB and a strong IGMF, which is consistent with the results in GR78 and Ref.\cite{aharonian92}.
When we enhance the strength of the IGMF to $B=10^{-8}\,$G case and $B=10^{-10}\,$G, synchrotron radiation start to cease the cascade process at high energies and the effective penetration distances approach to the mean free lengths of photons beyond $\sim 10^{7.5}\,$GeV and $\sim 10^{9.5}\,$GeV, respectively. We can also see in the presence of CRB, the effective penetration lengths become much smaller at EHE energy, as is discussed above. 
At higher redshifts, due to a higher CMB density, the effective lengths systematically shift downward. Also, the breaks in the curves due to CRB and IGMF shift to higher energy.

\begin{figure}[hptb]
\includegraphics[width=1.0\linewidth]{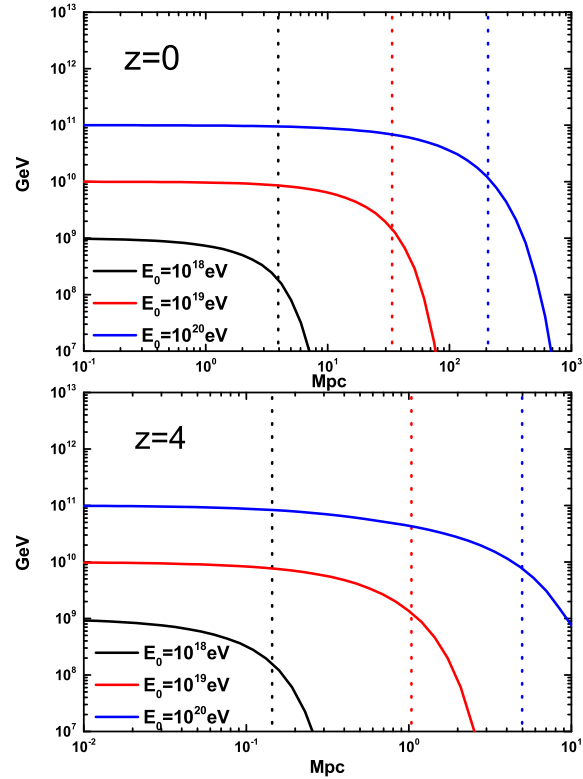}
\caption{\small{\label{f4}
The energy loss as the propagation distance at $z=0$ and $z=4$ in CMB gas when IGMF $B=10^{-12}\,\mathrm{G}$. The solid lines represent the averaged remaining energy with a different initial energy $E_0$ and are obtained by Monte Carlo simulation. The vertical dotted lines are used to indicate the corresponding effective penetration distances in Fig.~\ref{f3}.
}}
\end{figure}

To see how well the result of our analytic treatment approaches that of a more detailed numerical calculation, we utilize Monte Carlo simulation to trace the propagation of each leading particle initiated by a number of total $N=20000$ UHE photons with identical initial energy $E_0$. The energies of the leading particles decrease after undergoing interactions of EPP, CS, and TPP, as well as synchrotron loss if a strong IGMF is adopted. In the meanwhile,  particles entering the DPP and MPP channel do not generate leading particles, so the number of leading particles decreases as well.
We record the energy of each leading particle after each interaction and obtain the average energy of leading particle as a function of propagation distance by taking the average $\sum\limits_k {{E_k}(d)} /N$, with $E_k(d)$ denoting the energy of $k$-th leading particle after propagating a distance of $d$. In Fig.~\ref{f4}, we show the evolution of the typical energy of the leading particle with propagation distance by solid curves with different colors for different initial energy of $10^{18},\,10^{19},\,10^{20}\,\mathrm{eV}$ respectively. The effective penetration distances of UHE photons at these energies obtained by our analytic method are shown as vertical dashed lines with corresponding colors. We can see that, in most cases, particle's energy drops to about 20\% after propagating a distance equal to the corresponding effective penetration distances shown in Fig.~\ref{f3}. This validates that the effective penetration distance represents the distance after propagating which the total initial energy of the UHE particles are significantly lost. However, for photons of $E_0=10^{20}\,\mathrm{eV}$ injected at $z=4$ the effective penetration distance obtained by analytic method corresponds to the distance where $\lesssim 10\%$ of the initial energy is left. This is because that DPP becomes almost as important as EPP in this case, and hence a considerable fraction of interactions in $\gamma \to e$ enter DPP without producing leading particles, resulting in a faster decrease of the average energy of the leading particle.

\begin{figure}[hptb]
\includegraphics[width=1.0\linewidth]{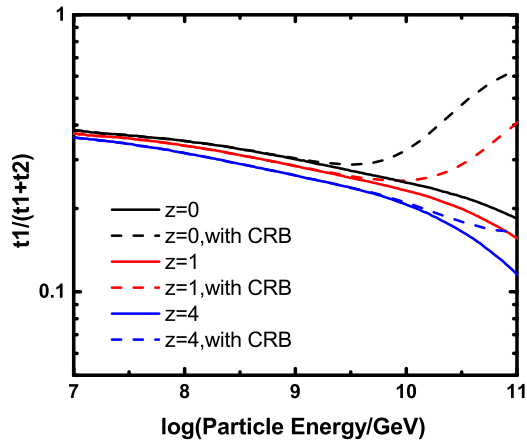}
\caption{\small{\label{f5}
The fraction of the time that particle propagates as a photon in IGMF $B=10^{-12}\,\mathrm{G}$  without CRB and with CRB at different redshifts.
}}
\end{figure}

An UHE particle will switch between a photon and an electron in the propagation. One may be interested in the distance traveled by the particle as a photon. The effective penetration distance of UHE particle as a photon can be estimated by
\begin{equation}
{\lambda _{eff,\gamma }} = c{t_1}\frac{E}{{\Delta {E_{tot}}}}.
\end{equation}
When DPP process is not important, ${E}/{\Delta {E_{tot}}}$ is roughly the number of the cycle $\gamma \to e \to \gamma$ that particle undergoes. Conversely, ${E}/{{\Delta {E_{tot}}}}$ is close to $1$ in the case that DPP is very important (e.g., when $E> 10^{21}\,$eV), and no leading particles will be produced. So the distance that an UHE particle travels as a photon is roughly the mean free length of DPP process, which is about $ct_1$ in this case.
We also present the fraction of the time that a particle propagates as a photon, i.e. $t_1/(t_1+t_2)$, as a function of the initial photon energy in a weak IGMF $B=10^{-12}\,\mathrm{G}$  in Fig.~\ref{f5}. This fraction would be suppressed by DPP cooling at high energy part in the case without CRB. As redshift increases, the influence of DPP extends to lower energy, due to a higher CMB temperature.
Besides, one may notice that an increase of the fraction appears at high energies when taking CRB into account. This is because, as can be seen from Fig.~\ref{f1} and Fig.~\ref{f2}, the decrease of the mean free path of CS is more significant than that of EPP in the presence of CRB. In other word, $t_1$ and $t_2$ both get smaller in the presence of CRB, but the decrease of $t_2$ is larger than that of $t_1$. As a result, the value of $t_1/(t_1+t_2)$ goes up. For a stronger IGMF, the value of  $t_1/(t_1+t_2)$ will be larger because stronger magnetic field will make electrons loss energies faster via synchrotron radiation but have no influence on photons.

\section{An extra component of cosmogenic neutrinos } \label{sec:neupro}
``Cosmogenic neutrino'' usually refers to the neutrinos decayed from charged pions which are generated in UHECR propagation. Neutral pions, which will decay into gamma rays, are also generated in the propagation of UHECRs, so an extra component of neutrinos will be born via the MPP process and the decays of muons ${\mu ^ + }({\mu ^ - }) \to {e^ + }{\nu _e}{{\bar \nu }_\mu }({e^ - }{{\bar \nu }_e}{\nu _\mu })$.

\begin{figure}[hptb]
\includegraphics[width=1.0\linewidth]{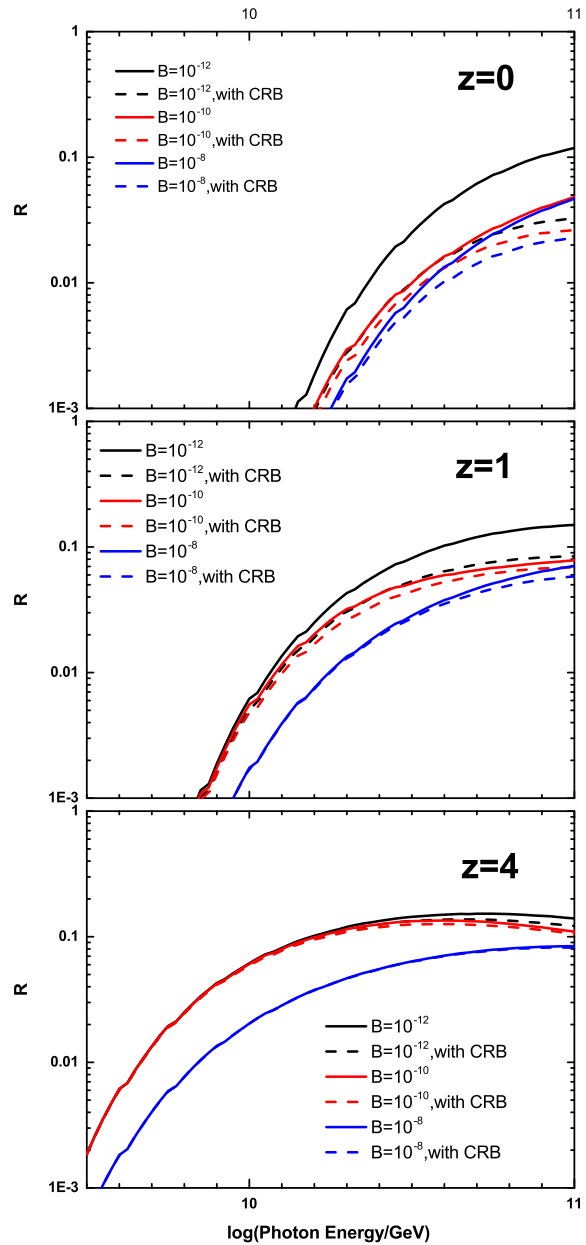}
\caption{\small{\label{f6}
The relative probability of going to MPP channel for an UHE photon at $z=0$, $z=1$ and $z=4$, respectively.
}}
\end{figure}

The energy loss of muon productions in one cycle of $\gamma \to e \to \gamma$ is
 ${\Delta {E_{mpp}}}=(dE/dt)_{mpp}t_1$.
 We then can find the ratio of the energy lost to MPP channel to the total energy lost in one cycle by
\begin{equation}
R = \frac{{\Delta {E_{mpp}}}}{{\Delta {E_{tot}}}}=\frac{{{{(dE/dt)}_{mpp}}}}{{{{\left\langle {dE/dt} \right\rangle }_{tot}}}}\frac{{{t_1}}}{{{t_1} + {t_2}}} .
\end{equation}
In Fig.~\ref{f6}, we can see that $R$ in weak IGMF is significantly larger than that in strong IGMF because the effective penetration distance get larger than the usual mean free length of UHE photon. Moreover, at high energy part, because of the relatively high threshold of MPP process ($\sqrt s  > 2{m_\mu } = 0.21\,\mathrm{GeV}$), interactions with CRB photons would lead to EPP rather than MPP, so $R$ would be smaller when one considers CRB. And the higher redshift would make the importance of IGMF and CRB down due to higher density of CMB photons.

Since two out of three secondary products of muon decay are neutrinos, the average neutrino production rate can be approximated by
\begin{equation}
 P_{\nu}=\frac{2}{3}\left(\frac{dE}{dt}\right)_{mpp}\frac{t_1}{t_1+t_2},
 \end{equation}
provided that three secondaries generated in the muon decay equally share the muon's energy. Note that the pion carries 1/5 of the proton's energy and it decays to two photons, each of which further produces two muons via MPP. If we assume all the secondaries produced in each above step share the energy of their parent particles, we can obtain that each of the extra cosmogenic neutrino takes about 1/60 of the parent proton's energy. On the other hand, a conventional cosmogenic neutrino takes about $1/20$ of the parent proton's energy, as a pion carries 1/5 of the proton's energy and each neutrino takes 1/4 of pion's energy if we also assume that the energy of a charged pion are shared equally by its four decay products. So, the energy of a conventional cosmogenic neutrino is about three times the energy of an extra cosmogenic neutrino from the same energy proton. However, eight neutrinos are generated from one neutral pion, contrast to three neutrinos from one charged pion. Assuming the produced ratio of $\pi^\pm$ to $\pi^0$ to be $1:1 $, the number of the extra cosmogenic neutrino is about $\frac{8}{3^\alpha}R$ times that of the conventional cosmogenic neutrinos at a fixed energy, given a primary UHECRs spectrum of $dN/dE_p  \propto E_p ^{-\alpha}$. So once we get the flux of the conventional neutrino produced at certain energy and certain redshift, we can get the flux of the extra neutrino with the ratio obtained above.

\begin{figure}[hptb]
\includegraphics[width=1.0\linewidth]{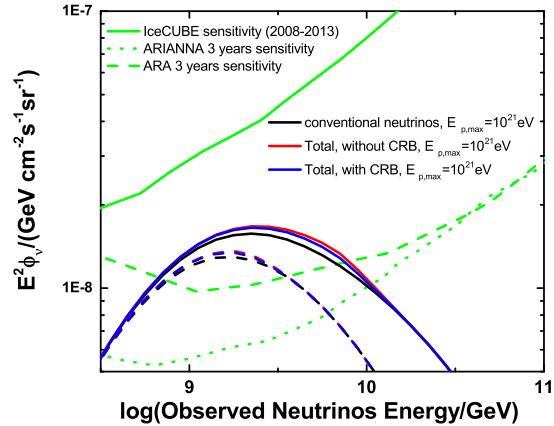}
\caption{\small{\label{f7}
The integral of all flavor neutrinos flux from $z=0$ to $z=5$ with IGMF $B=10^{-12}\,\mathrm{G}$. $E_{p,min}=10^{17} \,\mathrm{eV}$, $E_{p,max}=10^{21} \,\mathrm{eV}$ (for dashed lines, $E_{p,max}=10^{20.5} \,\mathrm{eV}$), $\int_{{E_{p,\min }}}^{{E_{p,\max }}} {{L_p}(E_p,z = 0)d{E_p} = {{10}^{44.5}}\,\mathrm{erg/{{Mpc}^3}/yr}}$. A proton spectrum index $\alpha=2$ at injection is assumed (i.e., $L_p\propto E^{-1}$). The black line represent the conventional cosmogenic neutrinos flux. The red and blue lines are the total fluxes including extra and conventional cosmogenic neutrinos for the cases without CRB and with CRB, respectively. The green lines denote the sensitivities of IceCUBE (2008-2013) and two future EeV neutrino experiments, ARA and ARIANNA (3 years) \cite{barwick15}.
}}
\end{figure}

\begin{figure}[hptb]
\includegraphics[width=1.0\linewidth]{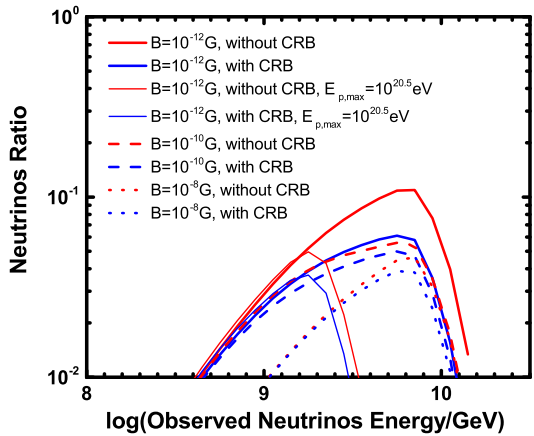}
\caption{\small{\label{f8}
The all flavor neutrinos ratio of extra neutrinos to conventional cosmogenic neutrinos as a function of observed energies of neutrinos. The thick lines are plotted using the same parameters as in Fig.~\ref{f7} except IGMF $B$ as a new parameter. Two thin lines (red and blue) are plotted if $E_{p,max}=10^{20.5} \,\mathrm{eV}$.
}}
\end{figure}

The conventional cosmogenic neutrino flux at the Earth can be given by integrating the contributions of individual UHECR sources at different cosmological epochs,
  \begin{equation}
{E_\nu }{\phi _\nu } = \frac{c}{4\pi }\int_0^{z_{\rm max}} \frac{1}{{1 + z}}f(E_\nu; E_p,z){L_p}(E_p,z)\frac{dE_p}{E_p} \frac{dt}{dz}dz.
  \end{equation}
where $f (E_\nu; E_p, z)$ is the observed neutrino spectrum ($E_\nu dN_\nu/dE_\nu$) produced during the propagation of a proton with energy $E_p$ injected at redshift $z$. This term is found by a Monte-Carlo code that was used and depicted in \cite{Liu16}, which is consistent with previous calculations of the conventional cosmogenic neutrinos \cite{yoshida93+}.  $dz/dt = {H_0}(1 + z){[{\Omega _M}{(1 + z)^3} + {\Omega _\Lambda }]^{1/2}}$, and we adopt $\Omega _M=0.3$, $\Omega _\Lambda=0.7$ and $H_0=70\,\mathrm{km/s/Mpc}$ in our calculations. $L_p(E_p,z)=L_p(E_p,z=0)S(z)$ is the UHECR differential emissivity at redshift $z$, and we assume $S(z)$ is redshift evolution of the UHECR emissivity which is assumed to follow that of the star formation rate \cite{yuk08}.
We present the conventional cosmogenic neutrinos and extra neutrinos in Fig.~\ref{f7}, and the ratio of extra neutrinos to conventional cosmogenic neutrinos in Fig.~\ref{f8}. If we fix the maximum energy of injected proton at $E_{p,\max} =10^{21}\,\mathrm{eV}$, which leads to a maximum photon energy $E_{\gamma,\max} \simeq 0.1E_{p,\max} = 10^{20}\,\mathrm{eV}$, by integrating  the redshift from $z=0$ to $z=5$, in a weak IGMF (e.g., $10^{-12}\,\mathrm{G}$), the neutrino flux from UHE photons can be up to about $11\%$ and $6\%$ of the conventional cosmogenic neutrino flux at the energy $E_{\nu}$ around a few EeV for the cases without CRB and with a constant CRB respectively. A strong magnetic field will lower this ratio. The magnetic field in the structured region (clusters and filaments with a size few Mpc) could be usually strong ($nG-\mu G$) \cite{gov04+}, and the mean free length of EPP is comparable with the size of structured region. So for the UHE photons generated at sources, the subsequent UHE electrons are produced in a strong magnetic field and loss their energy mainly through synchrotron radiation, which correspond to the case $B=10^{-8}\,\mathrm{G}$ in this paper and has been discussed in Ref.\cite{murase12}. In this work, we mainly focus on the UHE photons produced during the propagations of UHECRs in the cosmic background radiations. Basically, the energy loss lengths of UHECRs are $\sim$tens of Mpc or larger, which is typically much larger than the size of structured region. As a result, most of the UHE photons, in the situation we consider in this paper, tend to be produced in the region with a low magnetic field.
We also present the case with a lower maximum proton energy $E_{p,\rm max} =10^{20.5}\,\mathrm{eV}$ (i.e., $E_{\gamma, \rm max}\simeq 10^{19.5}\,$eV). As we can see, such a lower maximum proton energy significantly suppresses the production of the extra neutrinos. This is because of the high threshold of MPP process $\sqrt s  > 2{m_\mu } = 0.21$GeV. Given the typical energy of CMB photon to be $\epsilon=2.3 \times 10^{-4} (1+z)\,\mathrm{eV}$, we need the energy of UHE photon ${E_\gamma } >  {{{10}^{20}}\mathrm{eV}}/(1 + z)$ to enter the MPP channel. Thus, in the case of a low maximum proton energy, the produced pionic photons can only enter MPP channel when interacting with the high-energy tail of CMB photons at low redshits (e.g., $z <2$). We find the ratio of the extra neutrino flux to the conventional neutrino flux, integrated from $z=0$ to $z=5$, to be at most $\sim 5\%$ and $\sim 3.7\%$ for the cases without CRB and with a constant CRB respectively.

\begin{figure}[hptb]
\includegraphics[width=1.0\linewidth]{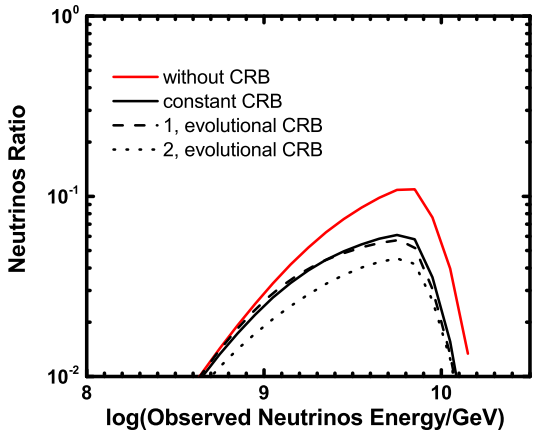}
\caption{\small{\label{f9}
The all flavor neutrinos ratio of extra neutrinos to conventional cosmogenic neutrinos in the cases with evolutional CRB when IGMF $B=10^{-12}\,\mathrm{G}$. There are two assumed forms of evolutional function $g(z)$: $(1)$ $g(z) = {(1 + z)^2}{e^{ - {{z}^2}}}$ ; And $(2)$ $g(z) = {(1 + z)^2}$ for $z<1$ and  $g(z) = 4$ for $z>1$.  The other parameters are same as used in Fig.~\ref{f7}.
}}
\end{figure}

We also consider possible evolutions of CRB  with redshift, although the evolution function is not clear so far. Ref.\cite{condon84} predicted an increase of the comoving density of radio sources with redshift at $z \lesssim 1$ and a quick drop at $z \gtrsim 1$, while some other analyses indicated that radio sources density may just slowly decrease at $z \gtrsim 2$ \cite{madau96+}. We assume CRB comoving density at $z$ to be $n(\varepsilon,z)=g(z)n(\varepsilon,z=0)$ , and check its effect on the extra cosmogenic neutrino with two different setups of $g(z)$: $(1)$ $g(z) = {(1 + z)^2}{e^{ - z^2}}$; $(2)$ $g(z) = {(1 + z)^2}$ for $z<1$ and  $g(z) = 4$ for $z>1$. The ratio of extra cosmogenic neutrinos to conventional cosmogenic neutrinos is shown in Fig.~\ref{f9}. Since the cross section of MPP peaks at $\sqrt{s}\simeq \sqrt{2E_\gamma \varepsilon} \simeq 0.4\rm GeV$ , neutrinos with lower energies tend to be generated at higher redshifts for a higher temperature of CMB as well as adiabatic cooling, and vice versa. So when adopting $g(z)$ of case 1, the amount of neutrino at high energy is lower than that in a constant CRB case because the CRB density is higher at low redshift, while amount of neutrino at low energy end increases and approaches that in no CRB case since a cut-off in $g(z)$ makes CRB density drops rapidly above $z=1$. Due to the difference between $g(z)$ of case $2$ and $g(z)$ of case $1$, the former one has a higher CRB density at $z>1$, so the amount of neutrinos at low energy in case 2 is much less than that in case 1.

The flavor ratio of conventional cosmogenic neutrinos is $f_{\nu_e}:f_{\nu_\mu}:f_{\nu_\tau}=1:2:0$ at production, resulting in a flavor ratio of $(f_{\nu_e}:f_{\nu_\mu}:f_{\nu_\tau})^{ob}=(0.93:1.05:1.02)^{ob}$ at Earth when considering oscillation with global best-fit mixing parameters \cite{gonza14}.
It might be interesting to note that the observed flavor ratio at Earth would be slightly different when considering such an extra component of cosmogenic neutrinos produced via muon pair production channel. Given the flavor ratio of this extra component to be $f_{\nu_e}:f_{\nu_\mu}:f_{\nu_\tau}=(1:1:0)^p$ at production, the flavor ratio at Earth is about $(1.10:0.94:0.96)^{ob}$ after oscillation. Assume the extra component contributes $10\%$ of the conventional component, the overall neutrino flavor ratio at production is $(1.045:1.955:0)^p$, which roughly yields an observed flavor ratio of $(0.945:1.04:1.015)^{ob}$ at Earth. However, this change in the flavor ratio is too slight to be recognized by the instrument in near feature. In astrophysical sources (e.g., GRBs), the change of flavor ratio has been discussed in Ref.\cite{razz06}.

\section{Discussions and Conclusions}\label{sec:d&c}
In this work, we studied the neutrino production in EM cascade initiated by UHE photons via the muon pair production process. To find out the neutrino production rate, we first presented a comprehensive study on the energy loss rate of an UHE photon in the propagation in intergalactic space.
Background photons that permeating the intergalactic space absorb the UHE photon and an electron/positron pair is produced. However, a new UHE photon, with energy only slightly less than the initial one, is possible to be regenerated via Compton scattering when considering the leading particle effect, and hence the UHE photon gets a second chance to produce neutrinos.
Regarding the energies carried away by the least energetic outgoing particle in the EPP and CS processes as energy loss, and also considering other possible processes (i.e., DPP, MPP, TPP and synchrotron radiation), we calculated the energy loss rate of the leading particle in the transformation chain $\gamma \to e \to \gamma$ of the UHE particle. We find that in the absence of cosmic radio background and strong IGMF, the energy of the initial photon may be carried to a much farther place than that the mean free path of the absorption by background photons. The effect of CRB and strong IGMF, which will suppress the regeneration of the UHE photon,  is also studied.

Then, based on the obtained energy loss rate, we calculated the neutrinos production rate through MPP and muon decay during the propagation of the UHE photons. Although the probability of entering MPP channel is small comparing with EPP and DPP process, the repeated regeneration of UHE photon provides a second chance of producing neutrinos. We found that an UHE photon at local universe can transfer up to $15\%$ of its initial energy to muon pairs, in the absence of CRB and strong IGMF. Note that without considering the neutrino production by the regenerated UHE photons, at most $\sim 10\%$ of the photon energy is expected to be converted to muon pairs given by \cite{li07}, even if neglecting DPP, which, however, should be important at the considered energy (so the percentage should be lower). Next, assuming a pure proton composition for UHECRs and equal amount of neutral pions and charged pions produced during the propagation of UHECRs, we found that the flux of neutrinos originated from neutral pions could reach  about a $10\%$ fraction of that of the conventional cosmogenic neutrinos produced in the decay of charged pion decay, by integrating the contribution of UHECRs injected over the whole universe. The accurate fraction depends on spectrum of the injected UHECRs and the redshift evolution of the injection rate. Considering CRB or assuming a strong IGMF will lower this fraction.

We note that one charged pion can also produce one UHE electron with energy $E_p/20$, which can enter this electromagnetic cascades chain and contribute the extra neutrinos production as well. The energy of neutrino produced by these UHE electrons are about 1/120 of the energy of the UHE proton.  Thus, for a given neutrino energy at Earth and an injection proton spectrum of $E^{-\alpha}$, the number of these electron-induced neutrinos is about $\frac{1}{2^{\alpha}}$ of that of the photon-induced neutrinos, given the ratio of generated $\pi^\pm$ to $\pi^0$ is $1:1 $. However, if IGMF is strong, these electrons would loss energy via synchrotron radiation quickly and have little chance to produce new UHE photons and subsequent neutrinos.

In addition to be generated during the propagation of UHECRs in intergalactic space, UHE photons could be generated at sources (e.g., AGN, GRB) via $pp-$collision with the gas therein or via photopion production with the radiation of the sources\cite{razz04,li07,murase09}, if UHECRs are accelerated there. UHE photons might also be produced by topological defects and decaying relic particles (see \cite{bahtt00,nishikov62+} for details). The regeneration of an UHE photon largely extends the observable range of UHE gamma--ray astronomy. Our work indicates a potential to detect EeV photons from nearby universe up to $\sim 10\,$Mpc or even farther if the magnetic field is not too strong and the radio background is weak. However, even if we assume the emissivity of UHE photon is the same as that of UHECRs, we estimate the flux fraction of UHE photon to UHECRs at Earth to be $\sim (10\,\rm Mpc/1\,\rm Gpc)=1\%$, given that $10\,$Mpc and $1\,$Gpc are the effective penetration length for 10\,EeV photon and proton respectively (see \cite{gel08} for a more comprehensive study). It is still hard to have a significant detection of these UHE photons since the current statistics of UHE photon detection by Auger results in a upper limit of 2.7\% for this fraction at 10\,EeV \cite{pao16}. On the other hand, detection of UHE photon from nearby transient sources, such as low-luminosity GRB and AGN giant flare, might be expected.\cite{murase09}. In future, with more accumulated data of current detectors and the next generation of detector such as JEM-EUSO \cite{JEUSO}, detection or non--detection of UHE photons will finally give some constraints on the strength of IGMF and CRB.


\appendix
\section{Interaction processes of UHE Photons}\label{photonsection}

\subsection{Electron Pair Production} \label{sectionepp}
During an UHE particle passing through an isotropically-distributed low energy photon gas with density $n(\varepsilon)$, the characteristic distance $L_c$ for the occurrence of a specified process, or the absorption probability per unit length $d\tau /dx$ can be written as
\begin{equation}
\begin{array}{lll}
    L_c^{ - 1} =  d\tau /dx & = & \frac{1}{2}\iint {{\sigma _{tot}}(1 - \cos\theta )}n(\varepsilon )d\varepsilon d\cos \theta  \\
   &= & \int {\left\langle {{\sigma}} \right\rangle n(\varepsilon )d\varepsilon },   \\ \label{CL}
\end{array}
\end{equation}
where $\theta$ is the collision angle between the velocity vectors of two incident particles in the laboratory frame, $\sigma_{tot}$ is the total cross section and $\left\langle {{\sigma}} \right\rangle$ is the averaged cross section on the collision angle.
For electron pair production process, the averaged cross section between the UHE photon with energy $E_\gamma$ and the low energy photon with energy $\varepsilon$ is \cite{aharonian83}
\begin{equation}
\begin{array}{lll}
  \left\langle {{\sigma _{epp}}} \right\rangle  =   \frac{{4\pi r_0^2}}{{{\omega_{\gamma} ^2}}} & |(\omega_{\gamma}  - 1 + \frac{1}{{2\omega_{\gamma} }} + \ln2 \sqrt{\omega_{\gamma}} )\ln (\sqrt \omega_{\gamma}  + \sqrt {\omega_{\gamma}  - 1} )    \\
    &  +  \frac{1}{8}{\ln ^2}\omega_{\gamma}  - \frac{1}{2}{\ln ^2}(\sqrt \omega_{\gamma}   + \sqrt {\omega_{\gamma}  - 1} )  \\
    & + \frac{1}{2}\ln2\ln \omega_{\gamma} -  \sqrt \omega_{\gamma}  \sqrt {\omega_{\gamma}  - 1} )|, \label{CSepp}
\end{array}
\end{equation}
where $\omega_{\gamma}  = \varepsilon {E_\gamma }$ in the units of the rest electron energy squared. Moreover, the produced differential electron and positron spectrum is suggested as
\begin{equation}
\begin{array}{ccc}
\frac{{dN}}{{d{E_e}d\varepsilon dt}} = &  \frac{{3{\sigma _T}c}}{{32}}\frac{1}{{{\varepsilon ^2}E_\gamma ^3}} \left[ \frac{{4{A^2}\ln (4\varepsilon {E_e}(A - {E_e})/A)}}{{{E_e}(A - {E_e})}} - 8\varepsilon A   \right.  \\
   & \left.   + \frac{{2(2\varepsilon A - 1){A^2}}}{{{E_e}(A - {E_e})}} - (1 - \frac{1}{{\varepsilon A}})\frac{{{A^4}}}{{E_e^2{{(A - {E_e})}^2}}} \right] n(\varepsilon ) \nonumber
\end{array}
\end{equation}
\begin{equation}
\frac{A}{2}\left( {1 - \sqrt {1 - \frac{1}{{\omega_{\gamma}}}} } \right) < {E_e} < \frac{A}{2}\left( {1 + \sqrt {1 - \frac{1}{{\omega_{\gamma}}}} } \right)
\label{eppspec}
\end{equation}
where $A=E_\gamma +\varepsilon$, and all energies are in units of the rest electron energy.

In the EPP process, two outgoing electron and positron, one is with energy larger and one is with energy less than $A/2$ ($A=E_\gamma + \varepsilon$), are created from photon-photon annihilation. We consider the energy loss rate of UHE particles corresponding to the transformation $\gamma \to e$ as same as GR78, through calculating the energies taken away by the least energetic outgoing particles,
\begin{equation}
 - \left( \frac{{d{E }}}{{dt}}\right)_{epp} = \int_0^\infty  {d\varepsilon \int_{{E_{e,\min }}}^{A/2} {d{E_e}{E_e}\left( {\frac{{dN}}{{d{E_e}d\varepsilon dt}}} \right)} }.
 \label{eppelr}
\end{equation}
Here $E_{e,\min }=\frac{A}{2}\left( {1 - \sqrt {1 - \frac{1}{{\omega_{\gamma}}}} } \right) $. One should note, in this paper, all energies are in units of the rest electron energy unless otherwise stated.

For the UHE photons with energies larger than $10^{15}\,\mathrm{eV}$ interacting with CMB, the EPP would enter the K-N regime ($\omega_{\gamma} \gg 1$), and also we have $A \simeq E_\gamma \gg 1$, $\varepsilon \ll 1$. One can obtain the averaged cross section is $3\sigma_T (\ln 2\sqrt{\omega_{\gamma}}-1) /2 \omega_{\gamma} $ approximately. The distribution of CMB photons can be described by
\begin{equation}
n(\varepsilon ) = \frac{m_e^3 c^3}{{{\pi ^2}{\hbar ^3}}}\frac{{{\varepsilon ^2}}}{{{e^{\varepsilon /kT}} - 1}},
\end{equation}
where $kT=kT_0(1+z)$ is the CMB peak energy in unit of rest electron energy and $z$ is the redshift. So the expression of absorption probability per unit length is \cite{gould67a,brown73a,gould78}
\begin{equation}
(d\tau /dx)_{epp} = \frac{{{\sigma _T}{{(kT)}^2}}}{{8{\Lambda ^3}{E_\gamma }}}\ln (0.47{E_\gamma }kT),
\label{absorpepp}
\end{equation}
where $\Lambda=\hbar /m_e c$ is the electron Compton wavelength. Substituting Equation (\ref{eppspec}) into (\ref{eppelr}), one can obtain approximately the energy loss rate of EPP in K-N limit is
\begin{equation}
 - \left( \frac{{d{E }}}{{dt}} \right)_{epp} \simeq \frac{{c{\sigma _T}{\left( {kT} \right)^2}}}{{64{\Lambda ^3} }},
 \label{eppelrcmb}
\end{equation}
which is independent of the energy of initial UHE photon.

\subsection{Double Electrons Pair Production}

For the higher-order process, DPP, based on detailed calculations \cite{cheng70a,cheng70b}, its cross section fast increases with center-of-momentum energy squared $s$ near the threshold and quickly approaches the asymptotic constant $\sigma_{dpp}  \approx 6.45 \times 10^{-30}\, \mathrm{{cm}^2}$. The evolutional cross section of DPP is described as $\sigma_{dpp}  \approx (1-4/\omega_{\gamma})^6 \times 6.45 \times 10^{-30}\, \mathrm{{cm}^2}$ by \cite{brown73b}. For simplicity, we ignore the DPP process between the CRB and UHE electrons because of too small cross section. Its absorption probability per unit length in high energy limits is ${(d\tau /dx)_{dpp}} = \int {{\sigma _{dpp}}n(\varepsilon )d\varepsilon } $.

Suggested by detailed numerical simulation \cite{demidov09}, the angular distribution of secondaries are mainly located at the forward and backward directions relative to the collision axis, and the initial energy of UHE photon is practically carried away by one of the pairs.  Since the electron and positron of the leading pair share the energy equally (i.e., no leading particle effect) and the energies of secondaries are significantly smaller than those of initial UHE photons, we consider the energies of UHE particles loss totally in DPP interactions. So one can write the energy loss rate of DPP for UHE particle as
\begin{equation}
-{\left( {\frac{{dE}}{{dt}}} \right)_{dpp}} = cE\int {{\sigma _{dpp}}n(\varepsilon )d\varepsilon },
\end{equation}
 and this energy loss rate in the CMB gas for the UHE photons with energies $\gtrsim 10^{15} \,\mathrm{eV}$ as
 \begin{equation}
 -{\left( {\frac{{dE}}{{dt}}} \right)_{dpp}} = cE \frac{{2\zeta (3){\sigma _{dpp}}(kT)^3}}{{{\pi ^2}\Lambda^3}},
 \label{dppelrcmb}
 \end{equation}
where $\zeta(3)=1.202$ is the Riemann zeta-function. The Equations (\ref{eppelrcmb}) and (\ref{dppelrcmb}) are equal at the critical energy $E_{cr,1}=4.5 \times 10^{19} /(1+z) \,\mathrm{eV}$, above which the energy loss rate of DPP would be dominant over EPP for the UHE photons. We also can obtain $E_{cr,2}=7.0 \times 10^{20} /(1+z) \,\mathrm{eV}$ by making $(d\tau /dx)_{epp}=(d\tau /dx)_{dpp}$ equal, above which the absorption probability of DPP would be larger than EPP's. Due to the leading effect in the EPP, the $E_{cr,1}$ is significantly smaller than $E_{cr,2}$.

\subsection{Muon Pair Production}

In addition to two processes mentioned above, the muon pair production (MPP) process $\gamma\gamma \to \mu^+ \mu ^- $ related to the neutrinos production should be interesting. The cross section of MPP can be easily achieved just replacing the electron mass by muon mass ($m_e \to m_\mu$) in Equation (\ref{CSepp}), and it is less by a factor $(m_e/m_\mu)^2$ at the peaks of two cross sections. Especially for a specific UHE photon with the energy above the threshold of MPP in the CMB gas, according to Equation (\ref{absorpepp}), the absorption probability of MPP can be easily written as
\begin{equation}
(d\tau /dx)_{mpp} = \frac{{{\sigma _T}{{(kT)}^2}}}{{8{\Lambda ^3}{E_\gamma }}}\ln \left( 0.47{E_\gamma }kT \frac{m_e^2}{m_\mu ^2} \right),
\end{equation}
where all energies and $kT$ are in unit of electron rest energy as well. $\sqrt s  > 2{m_\mu }c^2 $ would lead EPP enter the extreme deep K-N regime and the absorption probability of EPP can be $\kappa$ times larger than MPP only, here $\kappa \simeq (d\tau /dx)_{epp}/(d\tau /dx)_{mpp} \sim 10$. The UHE photons with a percentage about $x = 1 - \exp ( - {L_{\gamma}}/{L_{mpp}})$ would go to MPP channel, where ${L_\gamma} = {(L_{mpp}^{ - 1}+L_{epp}^{ - 1} + L_{dpp}^{ - 1})^{ - 1}}$ and $L_i=(d\tau /dx)_i^{-1}$ is the corresponding mean free length for $i$ process. For $\sqrt s  > 2{m_{{\pi ^ \pm }}}c^2 = 0.28\mathrm{GeV}$, the charged pions also can be generated. However, since the cross section of charged pions is smaller than MPP and the charged pions would decay to muons, we neglect this process.

The produced $\mu^\pm$ with lifetime ${t_\mu } = 2.1 \times {10^4}\,\mathrm{s}\,{E_{\mu ,\mathrm{EeV}}}$ would decay through ${\mu ^ + }({\mu ^ - }) \to {e^ + }{\nu _e}{{\bar \nu }_\mu }({e^ - }{{\bar \nu }_e}{\nu _\mu })$. The energy of UHE photon would loss totally because the cascades can not continue once MPP occurs. In CMB gas, the energy loss rate can be written as
\begin{equation}
-(dE /dt)_{mpp} = cE \frac{{{\sigma _T}{{(kT)}^2}}}{{8{\Lambda ^3}{E_\gamma }}}\ln \left( 0.47{E_\gamma }kT \frac{m_e^2}{m_\mu ^2} \right).
\end{equation}

\begin{figure}[hptb]
\includegraphics[width=1.0\linewidth]{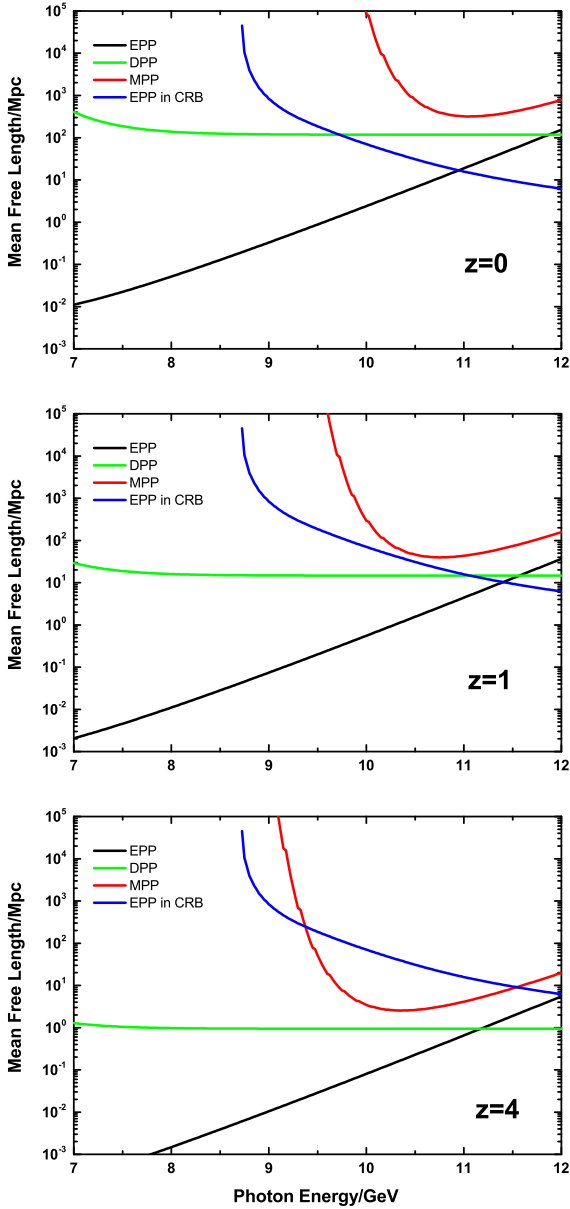}
\caption{{\label{f1}
The mean free lengths of EPP, DPP and MPP in CMB and CRB at $z=0$, $z=1$ and $z=4$ for UHE photons, respectively.
}}
\end{figure}

To sumarize, in Fig.~\ref{f1},  we show the mean free length of EPP, DPP and MPP in CMB photon field. We can see that in the case of no CRB, EPP process dominates over the other two processes below $7 \times 10^{20} /(1+z) \,\mathrm{eV}$. Above this critical energy, DPP is the most important process. The mean free length of EPP in CRB photon field is also presented. In the local Universe, the mean free length of EPP starts to dominate the transformation $\gamma \to e$ from $10^{20}\,$eV. But its influence becomes less important at higher redshift for a non-evolution CRB, as the number density of CMB increases as $(1+z)^3$. Note that CRB photon does not interact with UHE photons at the considered energy range via DPP and MPP due to the high threshold energy of DPP and MPP.

\section{Interaction processes of UHE Electrons}\label{electronsection}

\subsection{Compton scattering}\label{compscatter}
In the laboratory frame, the total cross section of Compton scattering of an electron with energy $E_e$ and a photon with energy $\varepsilon$ can be described by
\begin{equation}
\sigma_{cs} = \frac{3{\sigma _T}}{4s}\left[
(1 - 4/s - 8/s^)\ln(1 + s) + 1/2
 +8/s - \frac{1}{2(1 + s)^2}
\right]
\end{equation}
where $s = 2\varepsilon {E_e}(1 - {\beta _e}\cos \theta )$ is center-of-momentum energy squared, $\theta$ is the scattering angle and $\beta_e$ is the electron velocity. Its absorption probability per unit length is ${(d\tau /dx)_{cs}} =  \frac{1}{2}\iint {{\sigma _{cs}}(1 - \cos\theta )}n(\varepsilon )d\varepsilon d\cos \theta$. For calculating the energy loss rate of the UHE particle, as same as GR78, we consider the energy carried away by the least energetic outgoing particle as the lost energy. The energy range of scattered photon is from $\varepsilon  $ to ${E_e}{\Gamma _e}/(1 + {\Gamma _e})$ for the initial energies $\varepsilon$ and $E_e$ of the photon and electron, where $\Gamma_e=4\varepsilon E_e $ \cite{blumen70} (hereafter BG70). We classify the energy loss into two cases: ($1$) the main energy of initial UHE electron is taken away by scattered electron, $e \to e$, and ($2$) by scattered photon, $e \to \gamma$. We mark the case $1$ as "CS1" and the case $2$ as "CS2". For the CS1, the scattered photon is with the energy $\leqslant E_e/2$, and the main energy would be conserved in electron. The energy loss rate of CS1 can be computed through integrating over the scattered photon spectrum from $\varepsilon $ to $E_e/2$,
\begin{equation}
 - {\left( {\frac{{dE}}{{dt}}} \right)_{cs1}} =E_e \int_{{\varepsilon _{\min }}}^{{\varepsilon _{\max }}} {d\varepsilon } \int_{\varepsilon /{E_e}}^{1/2} {d{E_1}{E_1}\left( {\frac{{d{N_\gamma }}}{{dtd{E_1}d\varepsilon }}} \right)} ,
 \label{ics1}
\end{equation}
where the scattered photon spectrum ${\frac{{d{N_\gamma }}}{{dtd{E_1}d\varepsilon }}}$ is given by Equation ($2.48$) in BG70, where $E_1$ is the energy of the scattered photon in units of the initial electron energy. While for CS2, the scattered electron is with the energy $\leqslant E_e/2$, and the main energy would be converted to the scattered photon. Since the spectral shape of scattered electron and photon is symmetrical, so the energy loss rate of CS2 can be obtained by
\begin{equation}
\begin{split}
- {\left( {\frac{{dE}}{{dt}}} \right)_{cs2}} = E_e \int_{\varepsilon _{\min }}^{\varepsilon _{\max }}  {d\varepsilon } & \int_{1/2}^{{\Gamma _e}/(1 + {\Gamma _e})}  d{E_1}(1 - {E_1})\\
& \times \left( {\frac{{d{N_\gamma }}}{{dtd{E_1}d\varepsilon }}} \right),
\end{split}
\label{ics2}
\end{equation}
The absorption probabilities of CS1 and CS2,  $(d\tau/dx)_{cs1}$ and $(d\tau/dx)_{cs2}$, can be computed separately using a same manner as Equations (\ref{ics1}) and (\ref{ics2}) except the energy factors $E_eE_1$ and $E_e(1-E_1)$ should be rejected and two expressions should be divided by light speed $c$, respectively.

For the CMB photons, in the K-N regime, the analytic expressions of equations above can be given, for absorption probabilities (see GR78),
\begin{equation}
{\left( {\frac{{d\tau }}{{dx}}} \right)_{cs1}} = \frac{{{\sigma _T}{{(kT)}^2}}}{{8{\Lambda ^3}{E_e}}}(\ln2 + 3/8),
\end{equation}
\begin{equation}
{\left( {\frac{{d\tau }}{{dx}}} \right)_{cs2}} = \frac{{{\sigma _T}{{(kT)}^2}}}{{8{\Lambda ^3}{E_e}}}(\ln 2{E_e }kT + 9/8 - {C_E} - {C_l});
\end{equation}
While for energy loss rate,
\begin{equation}
 - {\left( {\frac{{dE}}{{dt}}} \right)_{cs1}} = \frac{{13c{\sigma _T}{{(kT)}^2}}}{{384{\Lambda ^3}}}
\end{equation}
\begin{equation}
 - {\left( {\frac{{dE}}{{dt}}} \right)_{cs2}} = \frac{{c{\sigma _T}{{(kT)}^2}}}{{16{\Lambda ^3}}}(\ln 2 - 5/12)
\end{equation}
where $C_E=0.5772$ is Euler's constant, and $C_l =0.5700$ (see BG70).

We treat the CS2 as the transformation $ e \to \gamma$ process, and the CS1 as a pure energy loss process. Suggested by GR78, the relative absorption probability ${(d\tau /dx)_{cs2}}/{(d\tau /dx)_{cs1}}$ increases as the energy of initial electron only logarithmically, and never gets very large. Also, both energy loss rates are constant and the  relative energy loss rate is ${(dE /dt)_{cs2}}/{(dE /dt)_{cs1}} \simeq 1.96$. We mark the total absorption probability of CS as $(d\tau/dx)_{cs}=(d\tau/dx)_{cs1}+(d\tau/dx)_{cs2}$ and the total energy loss rate as $(dE/dt)_{cs}=(dE/dt)_{cs1}+(dE/dt)_{cs2}$.


\begin{figure}[hptb]
\includegraphics[width=1.0\linewidth]{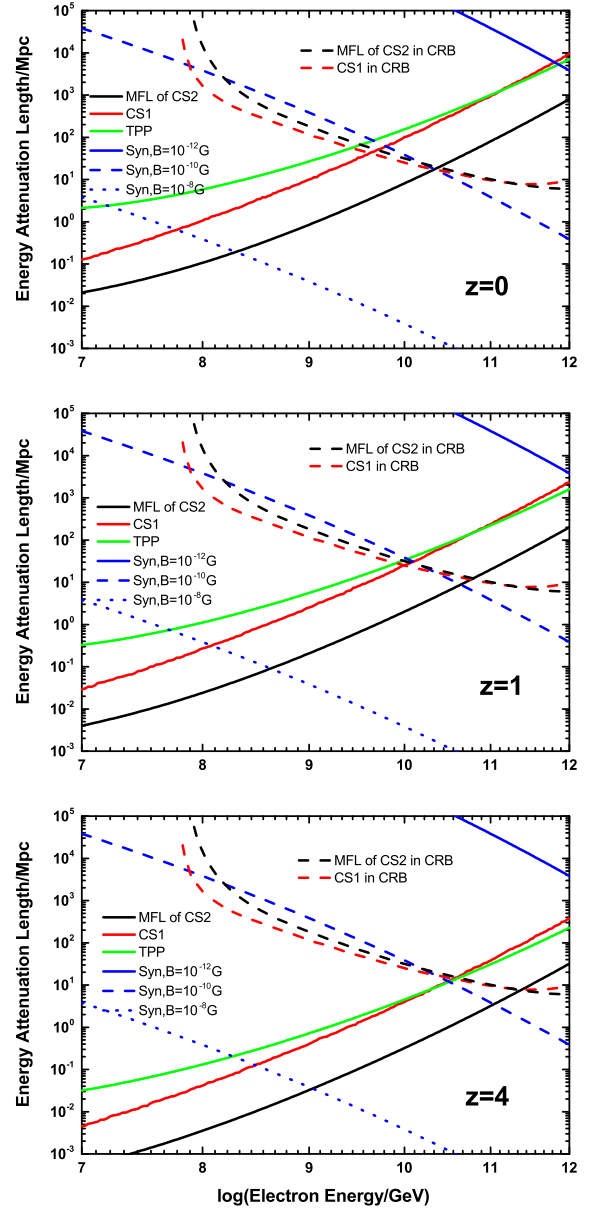}
\caption{\small{\label{f2}
The energy attenuation lengths of CS1, TPP and Synchrotron, and the mean free length of CS2 in CMB and CRB at $z=0$, $z=1$ and $z=4$ for UHE electrons, respectively.
}}
\end{figure}

\subsection{Triplet Production} \label{tppsection}

TPP process ($e\gamma \to e e^+ e^-$) is commonly ignored in astrophysical applications because of its extremely complicated differential cross section \cite{mork67+,mastich+}. The total TPP cross section was numerically evaluated by \cite{haug81}, who gives an analytic fit, at $\omega_e=\varepsilon E_e \gg 1$,
  \begin{equation}
  {\sigma _{tpp}} \simeq \frac{{3\alpha }}{{8\pi }}{\sigma _T}\left( {\frac{{28}}{9}\ln 2\omega_e - \frac{{218}}{{27}}} \right).
  \label{CStpp}
  \end{equation}
Basically, the absorption probability of TPP is considerably higher than CS's when $\omega_e \gtrsim \omega_{th,1}=220$. In terms of energy loss rate, it has been evaluated by some works (e.g., \cite{feenberg48+,mastich+}). Here we adopt a recent evaluation on the basis of precise numerical calculations given by \cite{anguelov99},
\begin{equation}
- {\left( {\frac{{dE}}{{dt}}} \right)_{tpp}} = \int {\frac{{0.145\alpha c{\sigma _T}}}{\varepsilon }\left( {{{\ln }^2}2\omega_e  - \frac{{218}}{{84}}} \right) n(\varepsilon )d\varepsilon },
\end{equation}

In CMB gas, according to Equations (\ref{CL}) and (\ref{CStpp}), the analytic expression of absorption probability is
\begin{equation}
{\left( {\frac{{d\tau }}{{dx}}} \right)_{tpp}} \simeq \frac{{7\alpha {\sigma _T}{{(kT)}^3}}}{{6{\pi ^3}{\Lambda ^3}}}2\zeta (3)(\ln2{E_e}kT - {C_E} - 1.1);
\end{equation}
While the energy loss rate is
\begin{equation}
 - {\left( {\frac{{dE}}{{dt}}} \right)_{tpp}} = \frac{{0.24\alpha c{\sigma _T}}}{{{\pi ^3}{\Lambda ^3}}}{(kT)^2}\left( {\ln 2{E_e}kT\ln 1.49{E_e}kT - 1.04} \right),
\end{equation}
 which gives us a threshold energy $\omega_{th,2} \sim 2.3 \times 10^7$ when equal to the energy loss rate of CS in case $2$. Such larger $\omega_{th,2}$ than $\omega_{th,1}$ implies the inelasticity of TPP is very small, i.e., one outgoing electron is leading extremely. So we ignore the TPP process when $\omega_e$ is small, because in this case, the CS process is much more dominant, especially for the interactions of UHE electrons with raido background.

\subsection{Synchrotron Radiation}

A well-known energy loss process for charged particles moving in moderate magnetic fields is the synchrotron radiation. In the chaotic magnetic fields, for the UHE electrons, the averaged energy loss rate is
\begin{equation}
 - {\left( {\frac{{dE}}{{dt}}} \right)_{syn}} = \frac{4}{3 m_e c^2}c{\sigma _T}E_e^2{U_B} ,
\end{equation}
where ${U_B} = \left\langle {{B^2}} \right\rangle /8\pi  $ is the averaged energy density of magnetic field, and except that $E_e$ and $kT$ are in the units of electron rest energy, all other items are in c.g.s. units.
During the propagation of UHE eletrons in the IGMF, to interrupt the transformation $e \to \gamma$ through CS1, one can obtain the energy range of electron
\begin{equation}
E_e^2 > \frac{{3\pi }}{4}\frac{{{{(kT)}^2}m_e c^2}}{{{\Lambda ^3}{B^2}}}\left( {\ln 2E_e kT - 0.022} \right) ,
\end{equation}
or
\begin{equation}
E_e \gtrsim 8.88 \times {10^{13}}\frac{{{{10}^{ - 10}} \,\mathrm{G}}}{B}(1 + z)
\end{equation}
when the energy attenuation length of synchrotron $cE/(dE/dt)_{syn}$  is larger than the mean free length $(1/(d\tau/dt)_{cs,2})$ of CS2.

To sumarize, in Fig.~\ref{f2}, we show the  mean free length of CS2, the attenuation length of CS1 and TPP in CMB field, and attenuation length of synchrotron radiation. CS2 is more important than CS1 and TPP at any energy and any redshift. As for synchrotron loss, when magnetic fields $B \gtrsim 4.5 \times {10^{-10}}\,\mathrm{G} (E_e/{{10}^{19}\,\mathrm{eV}} )^{-1}(1 + z)$, the synchrotron loss will significantly suppress the transformation $e \to \gamma$. We also show the mean free lengths of CS2 and the energy attenuation length of CS1 in the presence of CRB.  We can see CRB becomes more important than CMB for CS2 from the energy $\gtrsim 10^{19.5}\,$eV and for CS1 from $\lesssim 10^{19}\,$eV respectively in the local Universe, but its influence decreases with redshift, similar to that for EPP.

\begin{acknowledgments}
We thank Felix Aharonian for helpful comments and constructive suggestions about the interactions processes during the propagations of UHE photons. We also thank the anonymous referee for helpful comments and suggestions that have allowed us to improve the manuscript significantly. This work was supported by the National Basic Research Program (``973'' Program) of China under grant No. 2014CB845800 and the National Natural Science Foundation of China under grant No. 11573014 and grant No. 11273005.
\end{acknowledgments}

\clearpage

\end{document}